\documentstyle[12pt]{article}
\begin{document}
\title{
\hfill{\small TPBU-2-95}
\\
 \hfill{\small November 1995}
\\
\hfill{\small hep-th/9511121}
\\
\vspace*{0.5cm} \sc
On indices of the Dirac operator in a non-Fredholm case
\vspace*{0.3cm}}
\author{\sc Alexander Moroz\thanks{e-mail address :
{\tt am@th.ph.bham.ac.uk,  moroz@fzu.cz}}
\thanks{On leave from Institute
of Physics, Na Slovance 2, CZ-180 40 Praha 8, Czech Republic}
\vspace*{0.3cm}}
\date{
\protect\normalsize
\it School of Physics and Space Research, University of Birmingham,
Edgbaston, Birmingham B15 2TT, U. K.
}
\maketitle
\begin{center}
{\large\sc abstract}
\end{center}
 The Dirac Hamiltonian with the Aharonov-Bohm potential provides 
an example of a non-Fredholm operator for which all spectral 
asymmetry comes entirely from the continuous spectrum. In this 
case one finds that the use of standard definitions of the resolvent regularized, the heat kernel regularized, and the Witten indices 
misses the contribution coming from the continuous spectrum  and  
gives vanishing spectral asymmetry and axial anomaly. This behaviour 
in the case of the continuous spectrum seems to be general and its 
origin is discussed.

\vspace*{0.2cm}


\thispagestyle{empty}
\baselineskip 20pt
\newpage
\setcounter{page}{1}
\section{Introduction}
Let us consider an abstract Dirac Hamiltonian $H$ 
in $L^2($R${}^2)^2$ with
supersymmetry. It can be written in the form \cite{Th}
\begin{equation}
H=Q+M\tau,
\label{dsusy}
\end{equation}
where 
$M$ is a positive self-adjoint 
``mass'' operator, which commutes with $Q$ and $\tau$,
and $Q$,
\begin{equation}
Q=\left(
\begin{array}{cc}
0 & D^\dagger\\
D& 0
\end{array}
\right),\hspace*{1cm} \{Q,\tau\}=0,
\end{equation}
is a supercharge with respect to the involution $\tau$,
\begin{equation}
\tau=\left(
\begin{array}{cc}
1 & 0\\
0& -1
\end{array}
\right).
\end{equation}
The physical examples are provided by the Dirac Hamiltonian
with a scalar field, the Dirac Hamiltonian with and without anomalous
electric and magnetic moments in electromagnetic fields,
the Klein-Gordon equation in an external magnetic field, and Dirac type 
operators over Riemannian manifolds (see \cite{Th}, pp. 151-154
and references therein).

For a supercharge $Q$, the ``{\em Fredholm index}'' is defined as
\begin{equation}
\mbox{ind}\,Q\equiv
\mbox{dim Ker}\, D-
\mbox{dim Ker}\, D^\dagger=
\mbox{dim Ker}\, D^\dagger D-\mbox{dim Ker}\, DD^\dagger,
\label{fin}
\end{equation}
whenever this number exists.  Operators $D^\dagger D$ and $DD^\dagger$ are ``close'' in the sense that their spectra coincide except for 
the zero energy. In fact, it can be shown that their respective
restrictions on $(\mbox{Ker}\,D)^\perp$ and 
$(\mbox{Ker}\,D^\dagger)^\perp$  are unitary equivalent
(see \cite{Th}, p. 144). 
Therefore, in the case that the zero energy belongs to the point
spectrum, $\mbox{ind}\,Q$ is nothing but
\begin{equation}
\Sigma=\lim_{{\cal E}\downarrow 0}
\int^{\cal E}_{0_-} \sigma({\cal E})\, d{\cal E},
\label{sg}
\end{equation}
where $\sigma({\cal E})$ is the spectral asymmetry.
The latter can be defined as the difference between spectral
densities $\rho(E)$ for positive and negative energy continuums
(cf. \cite{AM89}),
\begin{equation}
\sigma({\cal E})\equiv \rho({\cal E})-\rho(-{\cal E}),
\end{equation}
where ${\cal E}=|E|\geq 0$, or, generally, as
\begin{eqnarray}
\sigma({\cal E}) &\equiv& -\frac{1}{\pi}\,\lim_{\epsilon\rightarrow 0}
\mbox{ImTr}\,\left(
\frac{1}{{\cal E}+i\epsilon -H} + \frac{1}{{\cal E}+i\epsilon +H}\right)
\nonumber\\
&=& -\frac{2{\cal E}}{\pi}\,\lim_{\epsilon\rightarrow 0}\mbox{ImTr}\,\left(
\frac{1}{{\cal E}^2+i\epsilon-D^\dagger D}+ 
\frac{1}{{\cal E}^2+i\epsilon-DD^\dagger}\right).
\label{sasy}
\end{eqnarray}

In the most physically important case, $Q$ is not Fredholm since 0
is not an isolated eigenvalue and it lies in the essential spectrum.
In this case, one must first define a regularized index.
The common choice is to use either the 
{\em ``resolvent regularized index''} or to define an index with 
the {\em ``heat-kernel regularization''} \cite{Th}.
To define the regularized indices,  we shall use the fact that
the Hilbert space ${\cal H}$ can be written as the direct sum
of the ``positive'', ${\cal H}_+$, and ``negative'', ${\cal H}_-$, 
subspaces with respect to involution $\tau$,
${\cal H}={\cal H}_+ \oplus {\cal H}_-$.
Accordingly, any operator $A$ in the Hilbert space can be written as 
\begin{equation}
A=\left(
\begin{array}{cc}
A_+ & A_{+-}\\
A_{-+}& A_-
\end{array}
\right).
\end{equation}
Then, according to Ref. \cite{Th}, we shall denote 
by {\em mtr} the diagonal sum (``the matrix trace'')  of $A$,
\begin{equation}
\mbox{mtr}\, A\equiv A_+ +A_-,
\label{mtr}
\end{equation}
and by {\em str} the ``supertrace'' of $A$,
\begin{equation}
\mbox{str}\, A\equiv \mbox{Tr}\,\mbox{mtr}\, A= \mbox{Tr}\,(A_+ +A_-),
\label{str}
\end{equation}
where Tr is the usual trace in the Hilbert space.
Now, if mtr$\,\tau(z-Q^2)^{-1}$ is trace class for some
$z\in C\backslash [0,\infty)$, the {\em ``resolvent regularized index''}, 
$\mbox{ind}_z Q$, is defined as \cite{Th},
\begin{eqnarray}
\lefteqn{\mbox{ind}_z Q\equiv z^{1/2} \mbox{str}\,\tau(z^{1/2}-Q)^{-1}
=z\,\mbox{str}\,\tau(z-Q^2)^{-1}}\nonumber\\
&&
= z\,\mbox{Tr}\,[(z-D^\dagger D)^{-1}-(z-DD^\dagger)^{-1}].
\label{dindz}
\end{eqnarray}
If $\exp \left(-Q^2t\right)$ is trace class for some $t>0$,
one defines $\mbox{ind}_t\,Q$, the index of $Q$ in the
{\em heat-kernel regularization}, as \cite{Th}
\begin{equation}
\mbox{ind}_t\,Q\equiv \mbox{str}\, \tau e^{-Q^2t}.
\label{indt}
\end{equation}
If $Q$ is not Fredholm, one cannot expect regularized indices  
$\mbox{ind}_zQ$ and $\mbox{ind}_t\,Q$ 
to be independent, respectively, of $z$ and of $t$.
Therefore, one defines a parameter independent index as
\begin{equation}
W(Q)\equiv\lim_{\stackrel{z\rightarrow 0}
{\scriptscriptstyle |\arg z|>\epsilon >0}}
\mbox{ind}_z Q =\lim_{t\rightarrow \infty}\mbox{ind}_t Q,
\label{indz}
\end{equation}
whenever either the first or the second limit exists.
Index $W(Q)$ is called the {\em Witten index} \cite{Th,Wi}.
This definition reduces to the notion of the Fredholm index 
whenever $Q$ is Fredholm \cite{Th}. The {\em axial anomaly} $A(Q)$ 
can be defined in  terms of $\mbox{ind}_zQ$ as \cite{Th}
\begin{equation}
A(Q)\equiv -\lim_{\stackrel{z\rightarrow \infty}
{\scriptscriptstyle |\arg z|>\epsilon >0}}
\mbox{ind}_z Q.
\label{daxan}
\end{equation}
 
In the case that the spectral asymmetry $\sigma({\cal E})$ comes 
from the point spectrum, the Witten index coincides with $\Sigma$. 
However, we shall show that  definition (\ref{indz}) of the Witten 
index is not equal to $\Sigma$ if the spectral asymmetry comes entirely 
from the continuous part of the spectrum. In particular, in the case
of the Dirac Hamiltonian in the presence of an Aharonov-Bohm 
potential, indices $\mbox{ind}_z Q$, $\mbox{ind}_t\,Q$,
and $W(Q)$, and consequently the
axial anomaly $A(Q)$ as defined by (\ref{daxan}), turn out to be zero.

\section{ Krein's spectral shift function}
\label{sec:sm}
In order to calculate the resolvent regularized index, $\mbox{ind}_z Q$,
we shall use Krein's formula \cite{F} which provides an
efficient way of calculating traces such as the trace in 
(\ref{dindz}).
Let us consider a pair of operators $T_1$ and $T_2$  in a
Hilbert space ${\cal H}$,
and define
\begin{equation}
G_1(z)=\frac{1}{z-T_1},\hspace*{1cm}G_2(z)=\frac{1}{z-T_2}\cdot
\end{equation}
According to Krein's formula \cite{F}, 
\begin{equation}
\mbox{Tr}\left[ G_1(z)-G_2(z)\right] =
\int_{-\infty}^\infty
\frac{\xi_{T_1T_2}(\lambda)}{(\lambda-z)^2}\,d\lambda,
\label{kform}
\end{equation}
where $\xi_{T_1T_2}(\lambda)$ is a bounded function, 
called Krein's {\em  spectral displacement operator} 
for the pair $T_1$ and $T_2$.
Krein's formula (\ref{kform})  is valid whenever
$T_1$ and $T_2$ are trace comparable, which means essentially that the
operation on the left hand side in (\ref{kform}) has meaning.
Both regularized indices (\ref{dindz}) and (\ref{indt}) can be expressed
in terms of Krein's spectral shift function $\xi$ \cite{Com},
\begin{equation}
\mbox{ind}_z Q
=\int_0^\infty \frac{z\xi(\lambda)}{(\lambda-z)^2}\,d\lambda,
\label{xiindz}
\end{equation}
\begin{equation}
\mbox{ind}_t\,Q=-t\int_0^\infty \xi(\lambda)e^{-\lambda t}\,d\lambda.
\label{xiindt}
\end{equation}
The Witten index and the axial anomaly are then given as \cite{Th}
\begin{equation}
W(Q)=-\xi(0)\hspace*{1cm}\mbox{and}\hspace*{1cm}{\cal A}=\xi(\infty).
\label{xiwa}
\end{equation}

In the case that  $T_1=H_0+V$ and $T_2=H_0$ are, 
respectively, perturbed and unperturbed
Hamiltonians, the spectral displacement
operator $\xi$ is  given as 
\begin{equation}
\xi_{T_1T_2}(\lambda)=\frac{i}{2\pi}\ln\,\det \mbox{{\bf S}}(\lambda),
\label{dxi}
\end{equation}
where {\bf S}$(\lambda)$ is the on-the-energy-shell S matrix \cite{BK}.
We have used formulas (\ref{kform}) and (\ref{dxi}) to calculate 
the change in the density
of states for various physical systems, including
the gravitational vortex \cite{AM89}, electromagnetic waves \cite{AMB},
the Pauli, the Schr\"{o}dinger \cite{AM20,AM30}, 
the Dirac, and the Klein-Gordon
Hamiltonians \cite{AM89} in the presence of the Aharonov-Bohm potential.

An important point is that validity of 
Krein's formula (\ref{kform}) together with
representation (\ref{dxi}) of Krein's  spectral displacement operator
$\xi(\lambda)$ is not restricted to a particular operator,
such as the Schr\"{o}dinger operator or the Dirac operator,
but can be applied to any pair of operators $T_1$ and $T_2$
(subject to the condition that they are trace comparable).
In the latter case, the formal S matrix can be defined in the same
way as the ``physical'' S matrix. Let $\psi_{T_1}$ and $\psi_{T_2}$
be asymptotic eigenstates  of  operators
 $T_1$ and  $T_2$. Then, for the purpose of
calculating the trace in (\ref{dindz}), the formal S matrix
associated with the pair $T_1$ and $T_2$ can be defined as
\begin{equation}
\psi_{T_1}=  \mbox{{\bf S}}\, \psi_{T_2}.
\end{equation}
For example, in the case that 
both $T_1$ and $T_2$  are rotationally symmetric,
the S matrix in the $l$th channel, where $l$ is the angular momentum, 
will be
\begin{equation}
\mbox{S}_l(\lambda)=\exp\left[2i \delta_{l;T_1/T_2}(\lambda)\right],
\end{equation}
where $\delta_{l;T_1/T_2}(\lambda)$  is the relative phase shift of 
the scattering
solution of $T_1$ with respect to the scattering solution
of $T_2$ in $l$th  channel.  Then, according to (\ref{dxi}),
\begin{equation}
\xi(\lambda)_{T_1T_2}=-\frac{1}{\pi}\sum_{l=-\infty}^\infty
\,\delta_{l;T_1/T_2}(\lambda).
\label{xis}
\end{equation}

In order to calculate the trace in (\ref{dindz}), the relevant pair
of operators is 
\begin{equation}
T_1=D^\dagger D,\hspace*{2cm} T_2=DD^\dagger.
\label{choice}
\end{equation}
Since spectra of operators $D^\dagger D$ and $DD^\dagger$
coincide except for the zero energy,
one expects that the relative S matrix for the pair of operators
$D^\dagger D$ and $DD^\dagger$
will be relatively simple. We shall confirm this expectation
by explicit calculations.

\section{The Dirac Hamiltonian in the Aharonov-Bohm potential}
Let us illustrate the above procedure for the case of the 
Dirac Hamiltonian in the presence of an Aharonov-Bohm (AB) potential
${\bf A}(r)$ \cite{AB} which, in the radial gauge, is given
by
\begin{equation}
A_r=0,\hspace*{1cm} A_\varphi=\frac{\Phi}{2\pi r}
=\frac{\alpha}{2\pi r}\,\Phi_0,
\label{abpot}
\end{equation}
where $\Phi=\alpha\,\Phi_0$ is the total flux through the flux tube
and  $\Phi_0$ is the flux quantum, $\Phi_0=hc/|e|$.
Let us write $\alpha=n+\eta$ where $n$ and $\eta$ 
are, respectively,  the integer and  the fractional part of $\alpha$.

After separation of variables in polar coordinates, the
Dirac Hamiltonian $H(A)$ reduces
to the direct sum, $H(A)=\oplus_l h_{m,l}$, of
radial channel operators $h_{m,l}$ in $L^2[(0,\infty), rdr]$,
\begin{equation}
h_{m,l}=\left[
\begin{array}{cc}
m &-i\left(\partial_r+\frac{\nu +1}{r}\right)\\
-i\left(\partial_r-\frac{\nu}{r}\right) & -m
\end{array}
\right],
\label{abdirac}
\end{equation}               
where $\nu=l+\alpha$ \cite{PG}. The supercharge in the $l$th channel
is 
\begin{equation}
Q_l=\left[
\begin{array}{cc}
0&-i\left(\partial_r+\frac{\nu +1}{r}\right)\\
-i\left(\partial_r-\frac{\nu}{r}\right) & 0
\end{array}
\right],
\label{q}
\end{equation} 
and 
\begin{eqnarray}
D^\dagger D &=& -\frac{d^2}{dr^2}-\frac{1}{r}\frac{d}{dr}+
\frac{\nu^2_+}{r^2} + g_m
\frac{\alpha}{r}\delta(r),\hspace*{1cm}\nu_+=l+\alpha,
\label{d+dd}\\
DD^\dagger & =& -\frac{d^2}{dr^2}-\frac{1}{r}\frac{d}{dr}+
\frac{\nu^2_-}{r^2} - g_m
\frac{\alpha}{r}\delta(r),\hspace*{1cm}\nu_-=l+1+\alpha,
\label{dd+d}
\end{eqnarray}
with $g_m=1$ \cite{Th,Hag}.

\subsection{Spectrum and scattering phase shifts}
The point spectrum of the Dirac Hamiltonian $H$ in the AB 
potential is empty \cite{AM89}.  There are no threshold modes
[$\equiv$ zero modes
in the case that $M=0$ in (\ref{dsusy})] in the spectrum. 
For a threshold state at $E=m$ to 
exist, the lower component $\chi_2$ of the Dirac spinor 
$\chi=(\chi_1,\chi_2)$ must be zero and the upper component $\chi_1$  
has to obey
\begin{equation}
\left(\partial_r -\frac{\nu}{r}\right)\chi_1(r)=0.
\end{equation}
Similarly, at the threshold $E=-m$, $\chi_1$ must vanish and
$\chi_2$ has to satisfy equation
\begin{equation}
\left(\partial_r +\frac{\nu+1}{r}\right)\chi_2(r)=0.
\end{equation}
These two equations can be easily integrated. Their respective
solutions are
\begin{equation}
\chi_1(r)=r^{\nu}\hspace*{1cm}\mbox{and} 
\hspace*{1cm} \chi_2(r)=r^{-(1+\nu)}.
\end{equation}
Obviously, neither $\chi_1$ nor $\chi_2$ are in $L^2[(0,\infty), rdr]$ 
for any $l$. If they are square integrable at infinity they are not 
so at the origin and {\em vice versa}.

Continuous spectrum of the Dirac Hamiltonian $H$ for $E>m$ in 
the AB potential is given in terms of Bessel functions 
\cite{PG,GJ},
\begin{equation}
\Psi_{{\cal E},l}=\chi(r) e^{il\varphi} e^{-i{\cal E}t/\hbar},
\label{dirspec}
\end{equation}
where
\begin{equation}
\chi(r)=\frac{1}{N}
\left(
\begin{array}{c}
\sqrt{{\cal E}+m}\,\, (\varepsilon_l)^l J_{\varepsilon_l\nu}(kr)\\
i\sqrt{{\cal E} -m}\,\, (\varepsilon_l)^{l+1} 
J_{\varepsilon_l(\nu+1)}(kr) e^{i\varphi}
\end{array}\right),
\label{dirspec1}
\end{equation}
$N$ is a normalization factor, and $\varepsilon_l=\pm 1$.
For $E=-{\cal E}<-m$, the scattering states are given by 
\begin{equation}
\Psi_{-{\cal E};l}(t,r,\varphi)=\Psi_{{\cal E};l}^*(t,r,
\varphi)|_{m\rightarrow -m}.
\label{-epsi}
\end{equation}
In what follows, ${\cal E}$ will stand for $|E|$.
The square integrability at the origin fixes the sign of 
$\varepsilon_l$ except for the channel $l= -n-1$ \cite{PG}. 
Except for the channel $l=-n-1$, phase shifts  of up and down
components of $\chi(r)$ are invariant under
the change of the sign of the energy, $E\rightarrow -E$, and are given by
\begin{equation}
\delta^{u}_l=\delta^{d}_{l}=\left\{
\begin{array}{rr}
-\pi\alpha,& l>-n-1,\\
\pi\alpha,&l<-n-1.
\end{array}\right.
\label{dconvshift}
\end{equation}

The crucial point for our discussion is
that two-component solutions of the massive Dirac equation have 
only {\em one degree of freedom} that is reflected in the equality
of up and down phase shifts \cite{GJ}.
The ambiguity in the channel $l=-n-1$, which
is relevant when different self-adjoint extensions of the Dirac Hamiltonian
in the Aharonov-Bohm potential are discussed 
\cite{AM89,AM20,AM30,PG,GJ,MT1}, 
does not play any role here, because, despite that the actual value 
of the phase shift depends
on the particular self-adjoint extension, up and down phase shifts in a 
given self-adjoint extension are always {em equal} \cite{Bo} (cf. Refs. 
\cite{AM89,PG,GJ,MT1}).
Therefore, in all channels the relative phase shift 
for the pair of operators $D^\dagger D$ and $DD^\dagger$ equals to zero,
\begin{equation}
\delta_{l;D^\dagger D/DD^\dagger} =
\left(\delta_{-n-1;D^\dagger D}-\delta_{-n-1;DD^\dagger}
\right)=0,
\end{equation}
and, consequently, the relative $S$ matrix is an identity operator,
\begin{equation}
\mbox{S}_l(\lambda)=\exp \left[ 2i\delta_{l;D^\dagger D/DD^\dagger}(\lambda)\right]
={\bf 1}.
\end{equation}
Then, according to (\ref{dxi}), Krein's spectral shift function
$\xi(\lambda)\equiv 0$, and by using (\ref{xiindz}) and (\ref{xiindt}) one has
\begin{equation}
\mbox{ind}_z\, Q=\mbox{ind}_t\,Q =0.
\label{resl}
\end{equation}
Similarly, by using definitions (\ref{indz}-\ref{daxan}),
or, relation (\ref{xiwa}), one has
\begin{equation}
W(Q)=A(Q)=0,
\end{equation}
which contradicts the result for the axial anomaly of the Dirac Hamiltonian
in the Aharonov-Bohm potential (see, for example, \cite{AM89,CO}).

\section{Discussion and conclusions}
\label{sec:con}
We have shown, by analyzing the example of the Dirac Hamiltonian
in the Aharonov-Bohm potential, that 
standard definitions (\ref{dindz}-\ref{daxan})
of the resolvent regularized, the heat kernel regularized, and
the Witten indices miss the contribution
coming from the continuous spectrum (cf. also Ref. \cite{EW}). 
This behaviour is not restricted 
to the special case of the Aharonov-Bohm potential but is valid
for a general potential. The crucial point, as it has been said above,
is that two-component solutions of the massive Dirac equation have 
only {\em one degree of freedom} that is reflected in the equality
of up and down phase shifts \cite{GJ}. Therefore, by using
Krein's formulas (\ref{xiindz}), (\ref{xiindt}), and (\ref{dxi}), 
the contribution of the continuous spectrum to indices 
will always be zero.
Indeed,  definitions (\ref{dindz}-\ref{daxan})
were tailored to count the difference in number of modes 
having, respectively, the upper and the lower component 
identically equal to zero.
The latter are the standard threshold (zero) modes.
However, both components of the Dirac spinor of a  scattering mode are 
not identically zero, and definitions (\ref{dindz}-\ref{daxan}) are not 
sensitive enough to reflect a nonzero spectral asymmetry in this case. 
The latter statement can be rephrased as follows: 
definitions (\ref{dindz}-\ref{daxan}) are not sensitive to the change
$E\rightarrow -E$ in the sign of the energy: in the example discussed here,
phase shifts in the critical channel $l=-n-1$ are not invariant
under the sign reversal of the energy, nevertheless
definitions (\ref{dindz}-\ref{daxan}) give a vanishing answer.
As a result, in the non-Fredholm and the noncompact cases one must use
definition (\ref{sasy}) for the spectral asymmetry 
in order to obtain nonzero axial anomaly and calculate Krein's 
displacement operator $\xi_{T_1T_2}$ for a corresponding choice of
operators $T_1$ and $T_2$ (see, for example, \cite{AM89}).
Otherwise, if one makes choice (\ref{choice}) for $T_1$ and $T_2$,
the definition of the spectral asymmetry \cite{Th} in terms
of Krein's displacement operator $\xi_{T_1T_2}$,
\begin{equation}
\sigma(H) =
\lim_{t\rightarrow 0}m\int_0^\infty \xi(\lambda)
\frac{d}{d\lambda}\frac{e^{-t(\lambda+m^2)}}{\sqrt{\lambda+m^2}}
= -\frac{m}{2}\int_0^\infty \xi(\lambda)(\lambda +m^2)^{-3/2}\,d\lambda,
\label{xisas}
\end{equation}
gives zero.

We hope that we have sufficiently demonstrated
the efficiency of Krein's formulas (\ref{kform}) and
(\ref{dxi}) for calculation of traces such as the trace in
(\ref{dindz}) in the case when 
scattering phase shifts are known,  and we think that 
Krein's formulas could be a useful substitute to the 
path integral calculation of indices of the Dirac Hamiltonian \cite{Al}.

I thank M. Cederwall for calling my attention to reference \cite{EW}.
This work was supported by EPSRC grant number GR/J35214.



\begin{thebibliography}{99}
\bibitem{Th}B. Thaller, {\em The Dirac Equation} 
(Springer, New York, 1992) Chap. 5.
\bibitem{AM89}A. Moroz, Phys. Lett. B {\bf 358}, 305 (1995);
Report IPNO-TH/89-94.
\bibitem{Wi}E. Witten, Nucl. Phys. B {\bf 188}, 513 (1981);
J. Diff. Geom. {\bf 17}, 661 (1982).
\bibitem{F}
J. M. Lifschitz, Usp. Matem. Nauk {\bf 7}, 170 (1952);
M. G. Krein, Matem. Sbornik {\bf 33}, 597 (1953);
J. Friedel, Nuovo Cimento Suppl. {\bf 7}, 287 (1958);
J. S. Faulkner, J. Phys. C\ {\bf 10}, 4661 
(1977).    
\bibitem{Com}In the case of $\mbox{ind}_t\,Q$, it is required that
$\mbox{ind}_t\,Q$ exists for some $t_0$. Then (\ref{xiindt})
is valid for all $t\geq t_0$ and the corresponding functions
$\xi$ coincide. See \cite{Th} for more details.
\bibitem{BK}M. L. Birman and M. G. Krein, Sov. Math.-Dokl.
{\bf 3}, 740 (1962).
\bibitem{AMB}
A. Moroz, Phys. Rev. B {\bf 51}, 2068 (1995).
\bibitem{AM20}A. Moroz, Mod. Phys. Lett. B{\bf 9}, 1407 (1995).
\bibitem{AM30}A. Moroz, Phys. Rev. A {\bf 53}, 669 (1996).
\bibitem{AB}
Y. Aharonov and D. Bohm, Phys. Rev.\ {\bf 115}, 485 (1959);
\bibitem{PG}P. de Sousa Gerbert, Phys. Rev.\ D {\bf 40}, 1346 (1989).
Since in the $l=-(n+1)$-th channel the orders
of the Bessel functions for the up/down component are respectively
1 and 0 when $\eta=0$, we have taken $\nu=\eta-1$ instead of
$\nu=-\eta$.
\bibitem{Hag}C. R. Hagen, Phys. Rev. Lett.\ {\bf 64}, 503 (1990).
\bibitem{GJ}P. de Sousa Gerbert and R. Jackiw, Commun. Math. Phys.\ 
{\bf 124}, 229 (1989).
\bibitem{MT1}C. Manuel and R. Tarrach, Phys. Lett. B{\bf 301}, 72 (1993).
\bibitem{Bo}Although our conclusions are valid for any linear combination
of the two possible solutions in the critical channel $l=n-1$,
this case is not relevant for our (supersymmetric) discussion. 
If one starts with a flux tube with a finite
radius $R$, then, to obtain the nontrivial linear combination in
the limit $R\rightarrow 0$, an attractive potential must be placed
inside the flux tube (see Refs. \cite{AM89,AM20,AM30,PG,Hag,MT1}), 
which is not the case considered here. 
Moreover, then the full Hilbert space contains a bound state which
explicitely breaks the supersymmetry \cite{AM89}.
\bibitem{CO}T. Jaroszewicz, Phys. Rev. D {\bf 34}, 3128 (1986);
R. Musto, L. O'Raifeartaigh, and A. Wipf, Phys. Lett. B {\bf 175}, 433 
(1986); A. Comtet and S. Ouvry, Phys. Lett. B {\bf 225}, 272 (1989).
\bibitem{EW}E. Weinberg, Nucl. Phys. B {\bf 203}, 445 (1982).
\bibitem{Al}L. Alvarez-Gaume, Commun. Math. Phys. {\bf 90}, 161 (1983);
D. Friedan and P. Windey, Nucl. Phys. B {\bf 235}, 395 (1984);
N. V. Borisov and K. N. Ilinski,  J. Sov. Math., 156 (1994).
\end{thebibliography}
\end{document}